# Self-doping processes between planes and chains in the metal-to-superconductor transition of $YBa_2Cu_3O_{6.9}$

M. Magnuson[1], T. Schmitt[2] V.N. Strocov[2], J. Schlappa[2,3],
A. S. Kalabukhov[4] and L.-C. Duda[5]

[1]Department of Physics, Chemistry and Biology, IFM, Thin Film Physics Division, Linköping University, SE-58183 Linköping, Sweden
[2]Paul Scherrer Institut, Swiss Light Source (SLS), CH-5232 Villigen PSI, Switzerland
[3]Institut Methoden und Instrumentierung der Forschung mit Synchrotronstrahlung (G-ISRR), Helmholtz-Zentrum Berlin für Materialien und Energie GmbH, D-12489 Berlin, Germany
[4]Quantum Devices Physics Group, Department of Microtechnology and Nanoscience-MC2, Chalmers University of Technology, SE412 96 Gothenburg, Sweden
[5]Department of Physics and Astronomy, Division of Molecular and Condensed Matter Physics, Uppsala University, Box 516, S-751 20 Uppsala, Sweden

## Abstract

The interplay between the quasi 1-dimensional CuO-chains and the 2-dimensional $CuO_2$ planes of $YBa_2Cu_3O_{6+x}$ (YBCO) has been in focus for a long time. Although the CuO-chains are known to be important as charge reservoirs that enable superconductivity for a range of oxygen doping levels in YBCO, the understanding of the dynamics of its temperature-driven metal-superconductor transition (MST) remains a challenge. We present a combined study using x-ray absorption spectroscopy and resonant inelastic x-ray scattering (RIXS) revealing how a reconstruction of the apical O(4)-derived interplanar orbitals during the MST of optimally doped YBCO leads to substantial hole-transfer from the chains into the planes, i.e. self-doping. Our ionic model calculations show that localized divalent charge-transfer configurations are expected to be abundant in the chains of YBCO. While these indeed appear in the RIXS spectra from YBCO in the normal, metallic, state, they are largely suppressed in the superconducting state and, instead, signatures of Cu trivalent charge-transfer configurations in the planes become enhanced. In the quest for understanding the fundamental mechanism for high-$T_c$-superconductivity (HTSC) in perovskite cuprate materials, the observation of such an interplanar self-doping process in YBCO opens a unique novel channel for studying the dynamics of HTSC.

Corresponding author: Martin.Magnuson@ifm.liu.se

## Introduction

One of the most intriguing problems that remain to be solved in correlated electron physics is finding the mechanism of high $T_c$-superconductivity (HTSC). The discovery of HTSC was made first for the cuprate perovskites that consist of planes of edge-sharing $CuO_2$ square plaquettes separated by layers containing rare-earth metals and oxygen. A transition to superconductivity below the critical temperature can be detected e.g. by electric conductivity measurements or by the response to an external magnetic field [1]. Fig. 1A shows the magnetic susceptibility of the optimally-doped compound $YBa_2Cu_3O_{6.9}$ as a function of temperature with a transition temperature of $T_c$ = 90.5 K. $YBa_2Cu_3O_{6+x}$ (YBCO) is a unique case for a superconducting cuprate because the planes of corner-sharing $CuO_2$ plaquettes are linked by additional layers consisting of one dimensional $CuO_3$ chains that are linked by apical oxygen, i.e. O(4) in Fig. 1B. The O(4) $2p_z$-orbitals hybridize with Cu(1) $3d_{y2-z2}$-orbitals in the chains and with Cu(2,3) $3d_{z2-r2}$-orbitals in the planes. The $CuO_3$ chains of YBCO can be more or less complete depending on the oxygen doping (oxygenation) level (x=[0,1]). Oxygenation leads to transfer of charge carriers (holes) between the chains and planes and thereby tunes the in-plane conductivity via the hole density of the $CuO_2$ planes [2]. Theoretically, the application of pressure [3] increases charge transfer between the planes and the chains. *Ab initio* local spin density functional calculations have also found a strong coupling between the valence fluctuations of the chain-Cu and the bridging O(4) ions in highly doped YBCO [4]. This is significant because antiferromagnetic spin fluctuations [5], phonons [6], and polarons [7] are all candidates that may have an influence on the Cooper pairing mechanism in the $CuO_2$ planes. It is also generally recognized that unconventional *d*-wave symmetry [8,9] and charge stripes may be involved. Recently, a pairing model based on magnetic interactions has been scrutinized [10] with resonant inelastic x-ray scattering (RIXS) by studying the spin excitation spectrum of YBCO at different doping levels.

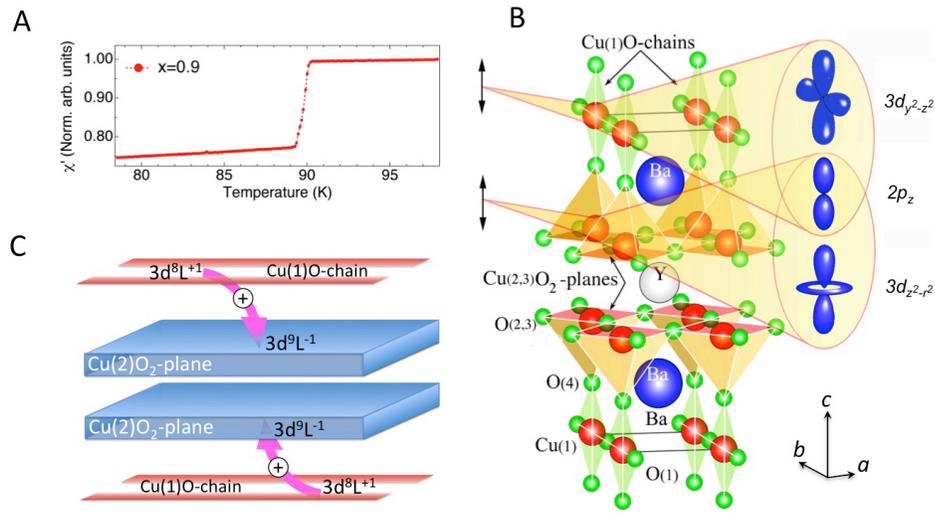

**Figure 1:** A) Magnetic susceptibility measurements of optimally doped $YBa_2Cu_3O_{6.9}$. B) Schematic illustration of the charge-transfer between chains and planes, taking place when YBCO is cooled through the MST. In the superconducting phase the $3d^8L^{+1}$ configuration at Cu(1)-sites in the chains is suppressed, while at Cu(2)-sites the $3d^9L^{-1}$ configuration is increased. The net effect is a self doping of the superconducting planes in the MST. C) The relevant orbitals for the self doping mechanism are the (partially empty) $3d_{3z2-r2}$ and $3d_{y2-z2}$ orbitals at the Cu(1) and Cu(2,3)-sites and the connecting apical oxygen O $2p_z$ orbitals. The first ones are preferentially excited when aligning the x-ray polarization vector along the *c*-direction (out-of-plane).

Charge-density modulations along the $CuO_3$ chains suggest that the surface CuO chain band develops a short-range charge density wave state as a result of a Peierls instability [11,12]. Discoveries in the band structure from highly surface sensitive angular resolved photoelectron spectroscopy (ARPES) [13,14] have been used to explain the in-plane anisotropy of YBCO [12,15] but the technique is less suitable to determine bulk properties. By contrast, x-ray absorption spectroscopy (XAS) is a powerful technique to study the character and density of hole states in the bulk. Early XAS studies of YBCO [15,16] have revealed large changes in the hole distribution between the $CuO_2$ planes and the $CuO_3$ chains as a function of oxygen doping. It would be desirable to also gain a detailed understanding of the *temperature dependent* XAS spectrum across the metal-superconducting transition (MST).

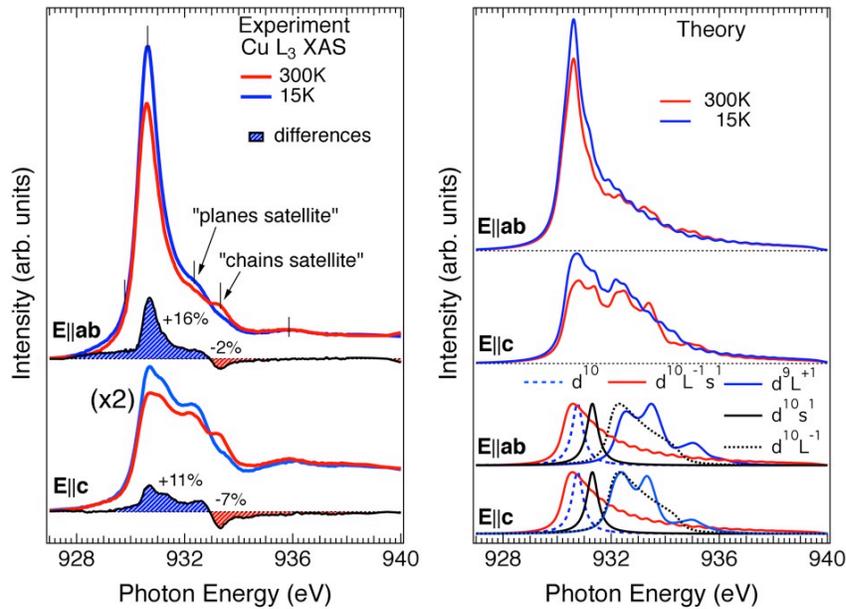

**Figure 2:** Left panel: experimental Cu 2p XAS-TFY spectra of optimally doped YBCO measured at 300 K (red lines), respectively 15 K (blue lines) and the difference spectra (15K – 300K). In the upper (lower) part of the panel, in-plane **E**||**ab** (out-of-plane, **E**||**c**), polarized x-rays have been used for excitation. The vertical ticks indicate the excitation energies for the RIXS measurements. Right panel: The corresponding simulated spectra are shown in the top part of the panel. They are obtained using fitted linear combinations of model calculations of spectra from core-hole excited $3d^{10}$, $3d^{10}L^{-1}4s^1$, $3d^9L^{+1}$, $3d^{10}4s^1$, and $3d^{10}L^{-1}$ final states, which are shown in the lower part of the panel.

In this work, we present evidence of the dynamic nature of the chain-plane charge transfer as part of the metal-to-superconductor transition of YBCO. It is accompanied by electronic reconstruction of the hybridized Cu(1)-O(4)-Cu(2)-orbitals (Fig. 1B). In effect, the $CuO_2$-planes undergo a self-doping due to a simultaneous configuration interaction between the Cu(1)- and the Cu(2)-sites (Fig. 1C). Thus our study reveals that self-doping is a crucial ingredient for metal-to-superconductor transition and HTSC in $YBa_2Cu_3O_{6.9}$.

## Results

Figure 2 (left panel) shows Cu $L_3$-XAS data from optimally doped YBCO, measured across the metal-superconducting transition (MST). The XAS and RIXS measurements were performed at the Advanced Resonant Spectroscopies (ADRESS) beamline [17] at the Swiss

Light Source (SLS), Paul Scherrer Institut, Switzerland, using the Super-Advanced X-ray Emission Spectrometer (SAXES) [18]. The XAS spectra have been measured in the bulk-sensitive total fluorescence yield (TFY) mode using polarization in-plane (**E**||**ab**) or out of plane (**E**||**c**), respectively. At the Cu 2p edge, the excitation process creates a 2p core hole at the metal site. While the main peak at 930.7 eV can be assigned to the $3d^9 \rightarrow 2p^5 3d^{10}$ transition, the satellites around 932.4 and 933.4 eV are due to metal-to-ligand charge transfer (MLCT) transitions on the Cu(2)- and Cu(1)-sites of the planes and the chains, respectively [19,20].

We find a strong temperature dependence of the spectral weight around the main peak but the most striking observation is that, regardless of the x-ray polarization direction, there are opposing intensity trends in the "chains satellite" and the "planes satellite" upon cooling. This suggests the intriguing possibility that there is a temperature driven redistribution of charge between the planes and the chains of YBCO. The spectral difference in percentages displayed in the figure are derived from the hatched areas relative to the total area (taking into account the atomic step between $L_3$ and $L_2$) under the room temperature spectra. The difference highlights the opposing behavior of the satellites in the MST. Formally, Cu ions in $YBa_2Cu_3O_{6.9}$ have a charge state of +2.3. For an intermediate mixed valence system, such as YBCO, one can expect that the ground-state wave function configurations of YBCO can fluctuate [4] between different Cu-site configurations and give rise to the transitions listed in Table I, where $L^{+1}$($L^{-1}$) denotes an additional (a missing) electron in the oxygen ligand band and such configurations are the effect of MLCT (see Supplementary).

At the top of the right panel of Fig. 2, we show simulated Cu $L_3$-spectra created by using linear combinations of the pure-configuration spectra at the bottom of the right panel. Since a full theoretical description of the Cu L-XAS spectrum of YBCO has proven to be too complex for state-of-the-art ab initio methods we employ an ionic model approach to project the charge-transfer contributions from essential configurations. Our calculations show that the MLCT transitions $3d^9 \rightarrow 2p^5 3d^9 L^{+1}$ deriving from formally divalent copper sites must heavily contribute to the spectral weight of the two satellites (reproduced by the double-peaked blue traces in Fig. 2). However, the opposing temperature trends in the satellites suggest the existence of additional contributions besides those from divalent copper sites. Thus the theoretical spectrum belonging to $3d^9 L^{+1} \rightarrow 2p^5 3d^{10} L^{+1}$ and $3d^9 L^{-1} \rightarrow 2p^5 3d^{10} L^{-1}$ (dashed blacked traces in Fig. 2) peaking only at the first satellite, indicates that monovalent and/or trivalent copper sites may contribute to the charge redistribution between the chains and planes. In order to distinguish between contributions from different configurations and to learn more about the nature of the states involved, we have performed incident-energy and -polarization dependent RIXS, in particular to study the transitions belonging to the satellite peaks.

**Table I:** Cu transitions in different configurations

| Transition | Ground State | Core-hole Excited State |
|---|---|---|
| Double hole/ZRS: | $3d^8 + 3d^9 L^{-1}$ | $2p^5 3d^9 + 2p^5 3d^{10} L^{-1}$ |
| Divalent Cu sites: | $3d^9 + 3d^{10} L^{-1} + 3d^8 L^{+1}$ | $2p^5 3d^{10} + 2p^5 3d^{10} L^{-1} 4s^1 + 2p^5 3d^9 L^{+1}$ |
| Monovalent Cu: | $3d^{10} + 3d^9 L^{+1}$ | $2p^5 3d^{10} 4s^1 + 2p^5 3d^{10} L^{+1}$ |

Figure 3 (left panel) shows an overview of the polarization dependent Cu $L_3$ RIXS spectra of optimally doped YBCO measured in the normal state (300 K) and in the superconducting state (15 K), respectively. The RIXS spectra were recorded at a 90°-scattering geometry with a grazing incidence angle of 20° with respect to the sample surface, i.e. for the geometry denoted by **E**||**c** there is an angle of 20° between the

polarization of the incident x-rays and the crystal **c**-axis. The RIXS spectra have been normalized to the incident x-ray flux and plotted relative to the incident x-ray energy, i.e. plotted on an energy loss scale. For convenience when comparing spectral shapes, we have applied scaling factors as indicated to spectral pairs of identical geometry.

Excitation close to the $L_3$ resonance (929.8 eV and 930.7 eV) yields spectra dominated by *dd*-excitations around -1.5 eV that are localized on divalent Cu orbitals. Moreover, magnetic and lattice excitations (here para-magnons and phonons, respectively) at energy losses within a few hundred meV are prominent at this excitation energy [10, 21]. Particularly, out-of-plane polarized components are found to be temperature sensitive (see also Supplemental Materials). Intriguingly, already early on, specific phonon modes have been suggested to be of crucial importance for HTSC of cuprates, e.g. by facilitating interplane pairing. Indeed, the $A_g$ Cu(1)-O(4) stretching vibration has been found to have anomalous strength in infrared reflectivity [22] and could be coupled to charge transfer [23]. Together with structural lattice instabilities in YBCO, the observation of *transverse modes* sensitive to the MST of YBCO is possible evidence for unconventional, e.g. polaron-mediated, superconductivity. Below, we show that YBCO indeed undergoes a massive out-of-plane electronic reconstruction in the MST as well.

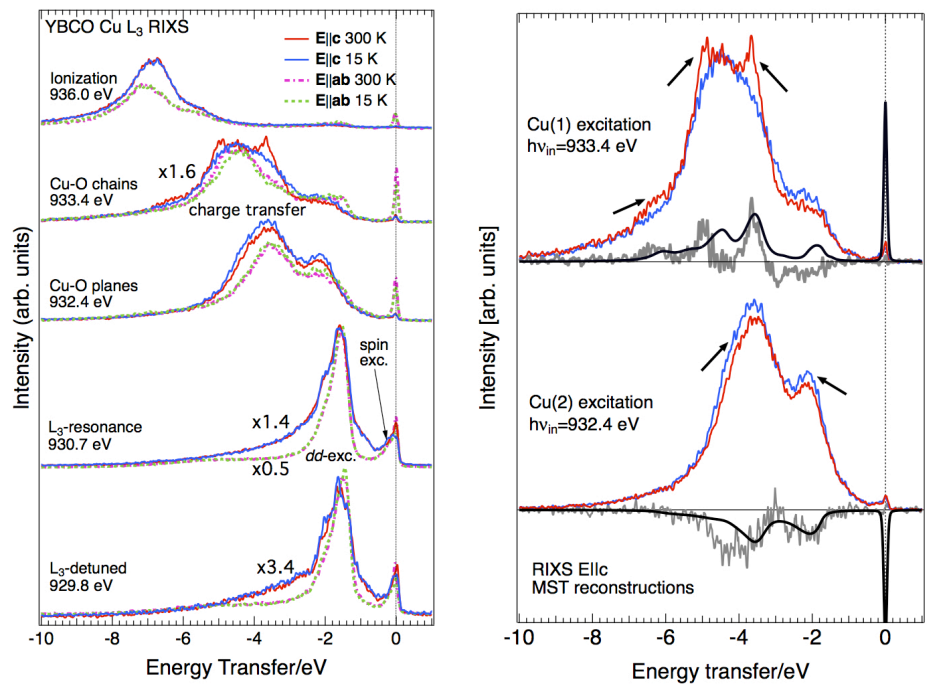

**Figure 3:** Left panel: Energy and polarization dependent Cu $L_3$ RIXS spectra (in-plane, **E**∥**ab** and out of plane, **E**∥**c**) of optimally doped YBCO measured in the normal state (300 K) and in the superconducting state (15 K), respectively. Spectral pairs are normalized to the incident photon flux. The arrow indicates the spin excitation peak (paramagnons) with -0.21 eV energy loss in the $L_3$ spectrum at 930.7 eV incident energy. Right panel: RIXS spectra excited at 933.4 eV ("chain satellite") (top) and at 932.4 eV (ZRS satellite) (bottom). The red (blue) lines are spectra recorded at 300 K (at 15 K). The grey lines are the respective experimental difference spectra (300K – 15K). The black lines are calculated spectra with enhanced configuration $3d^9L^{-1}$ for excitation at 932.4 eV and $3d^8L^{+1}$ for excitation at 933.4 eV (see Supplementary Materials for details).

RIXS excitation at the XAS satellite peaks are dominated by MLCT excitations and as such contain unique information about Cu-O bonding properties. Figure 3 (right panel) shows a close-up of the spectra excited at 932.4 eV (planes) and at 933.4 eV (chains). The general structure of the spectra are typical for a transition metal oxide, consisting of a main peak at higher energy loss that reflect states derived from pure O 2p-states and a shoulder at lower energy loss from oxygen states hybridized with the Cu 3d-band. The selfdoping effect across the MST observed in the Cu $L_3$-XAS also manifests itself in RIXS as a swap of sign in the difference spectra (gray traces in Fig. 3, right panel) when the exciation energy is changed from the plane satellite to the chain satellite – consistent to observations made with Cu $L_3$-XAS data. Dramatically, in addition to this, the MLCT excitations of the chain-Cu exhibit sharp, localized states of out-of-plane character that exist above the MST, yet are absent at low temperature. Recently, such structures in the O(4) $2p_z$-states have been theoretically predicted by first principal calculations [24]. There also exists good agreement with the expected polarization dependence of the O(4) 2p-DOS. Interestingly, we observe that the RIXS spectra show much greater temperature dependence for out-of-plane polarization (**E**||**c**) than in-plane polarization (**E**||**ab**), which correlates with the observed decrease of an out-of-plane lattice mode (see above). Thus we observe a reconstruction of orbitals that are aligned perpendicular to the planes and chains and are hybridized with apical oxygen O(4) $p_z$-orbitals. Propensity of these orbitals to reconstruct can also be inferred from a recent study at YBCO/manganite heterostructure interfaces [24].

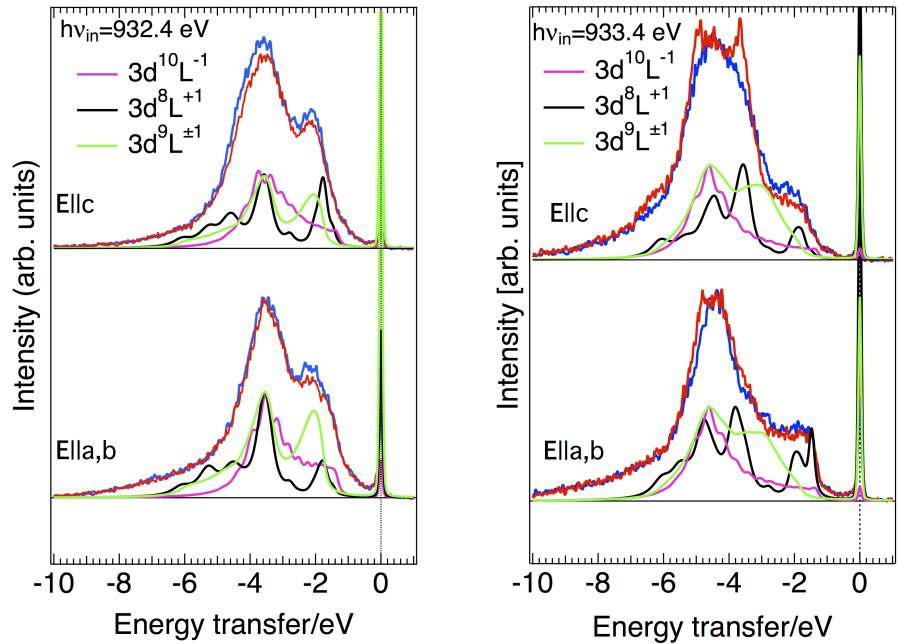

**Figure 4:** RIXS spectra excited at 932.4 eV (933.4 eV) in left (right) panel. Red (blue) lines are spectra recorded at room temperature (at 15 K). Corresponding model calculations using the indicated ground state configurations: *$3d^{10}L^{-1}$* (magenta), *$3d^8L^{+1}$* (black), and *$3d^9L^{±1}$* (green). In the lower (upper) part of the panel, in-plane **E**||**ab** (out-of-plane, **E**||**c**), polarized x-rays have been used for excitation.

## Discussion

Figure 4 shows a close up of the satellite excited RIXS spectra in comparison to calculations. The spectral shapes are well reproduced using our ligand field model calculations that account for the large contribution for MLCT excitations at energy losses higher than 3 eV, as well as the weaker, lower lying *dd*-excitations around 1.5 eV energy loss. In order to analyze this orbital reconstruction of the Cu(1)-O(4) bond we performed model calculations using the same ground state configurations as used for the XAS spectra. Particularly, the loss of spectral weight of the localized structures suggests that certain configurations become suppressed by cooling YBCO and it would be elucidating to identify which ones. Even though the raw RIXS spectra result from a superposition of several configurations listed in Table I, we find that the major contribution to the difference spectra (i.e. the electronic reconstruction) can be traced back to a single configuration at each of the incident energies. In Fig. 3 (right panel) we have superimposed calculated spectra (black traces) from configurations that display the best correspondence to the respective experimental temperature dependent difference spectra (gray traces). At the first XAS satellite peak, the RIXS differences correlate closest with those from theoretical spectra of $3d^9L^{-1}$ configuration, i.e. double hole/ZRS. Correspondingly, for excitation at the second XAS satellite peak, we find that the distinct sharp-peaked structure of the difference can only be reproduced by including a substantial contribution from the $d^8L^{+1}$ configuration. This involves formally divalent Cu ions that have transferred an electron into the ligand band. In other words, the net charge transfer that occurs in the MST is accounted for by the extra holes from the $d^8L^{+1}$ configuration of the chains that are pushed to the planes at low temperature and lead to additional $d^9L^{-1}$ double holes/ZRS (Fig. 2) [25]. Even though our calculations are based on a quasi localized theoretical approximation for the complex multi-electron state in YBCO, this already can consistently describe the overall intensity increase (reduction) of the corresponding RIXS spectra excited at the first (second) satellite peak (Fig. 2, 932.4 eV and 933.4 eV, respectively) as well as their basic spectral shapes [26].

In conclusion, we have studied the temperature-driven MST of YBCO and observed dramatic changes in its electronic structure. We find that the MST is accompanied by self-doping of the superconducting planes via interplanar charge transfer from the chains, possibly mediated by specific phonon modes. While YBCO in its normal, metallic state shows an abundance of quasi-localized Cu $3d^8L^{+1}$ configurations in the chains they are suppressed below the MST. Concurrently, this reconstruction of the interplanar Cu-O orbitals leads to an increase of the Cu $3d^9L^{-1}$ configuration in the planes. We expect that the study of charge-transfer excitations by polarization and temperature-dependent RIXS will prove extremely fruitful for gaining insight to the MST of a wide range of cuprate superconductors as well as, strongly correlated materials, in general.

## Methods
### Samples and experiments
Thin films of $YBa_2Cu_3O_{6+x}$ (YBCO) were grown by pulsed laser deposition on (100) $SrTiO_3$ 5x5x0.5 $mm^3$ single crystal substrates using 25 nm thick $CeO_2$ buffer layers deposited using RF sputtering at a temperature of 780 °C. High-resolution XAS and RIXS experiments were performed at the Advanced Resonant Spectroscopies (ADRESS) beamline [17] at the Swiss Light Source (SLS), Paul Scherrer Institut, Switzerland, using the Super-Advanced X-ray Emission Spectrometer (SAXES) [18]. A photon flux of $10^{13}$ photons/sec/0.01% bandwidth was focused to a spot size below 5x55 μm vertically and horizontally (VxH). The Cu $L_3$ edge XAS measurements were made in the bulk-sensitive total fluorescence yield (TFY) mode with an energy resolution of 100 meV. For the RIXS measurements at the Cu $L_3$ edge (930 eV), the combined energy resolution was 120 meV. The single-impurity Anderson model (SIAM) with full multiplet effects was applied to describe the system (see Supplementary).

## Acknowledgements
We would like to thank the staff at SLS for experimental support. This work was financially supported in part by the Swedish Research Council (VR), the Royal Swedish Academy (KVA) and MyFab-Micro and Nanofabrication infrastructure. This work was performed at the ADRESS beamline of the Swiss Light Source at the Paul Scherrer Institut, Switzerland.

## Author contributions
M. M. conceived the project. M. M., T. S., V. N. S. and J. S. carried out the experiments. A. S. K. fabricated samples and conducted sample characterization. M. M. conducted the calculations. M. M. and L. C. D. analyzed experimental and theoretical data and wrote the manuscript with input from all other authors.

## Additional information
Supplementary information accompanies this paper on http://www.nature.com/srep
Competing financial interests: The authors declare no competing financial interests for this system.